\documentclass[lettersize,journal]{IEEEtran}

\usepackage{cite}
\usepackage{amsmath,amssymb,amsfonts}
\usepackage{algorithmic}
\usepackage{graphicx}
\usepackage{textcomp}

\usepackage{amssymb}
\usepackage{graphicx}
\usepackage{url}
\usepackage{float}
\usepackage{amsfonts}
\usepackage{amsmath}
\usepackage{nccmath}
\usepackage{adjustbox}
\usepackage{multirow}
\usepackage{eurosym}
\usepackage{mathtools}
\usepackage{textcomp}
\usepackage{scrextend}
\usepackage{verbatim}
\usepackage{soul} 
\usepackage{fancyhdr,lipsum}
\usepackage{enumitem}

\usepackage{amsmath}
\usepackage{amsfonts}
\usepackage{mathtools}
\usepackage{physics}
\usepackage{dsfont}
\usepackage{verbatim}
\usepackage{color}
\usepackage[dvipsnames]{xcolor}
\usepackage{soul}
\usepackage[utf8]{inputenc}
\usepackage{textgreek}
\usepackage{amssymb}
\usepackage{multirow}

\usepackage{balance}
\usepackage{graphicx}
\usepackage{subfigure}
\usepackage{url}
\usepackage{nccmath}
\usepackage{adjustbox}
\usepackage{multirow}
\usepackage{eurosym}

\usepackage{cite}
\usepackage{amsmath,amssymb,amsfonts}
\usepackage{algorithmic}
\usepackage{graphicx}
\usepackage{textcomp}
\usepackage{xcolor}

\usepackage{amsmath,amsfonts}
\usepackage{algorithmic}
\usepackage{algorithm}
\usepackage{array}
\usepackage[caption=false,font=normalsize,labelfont=sf,textfont=sf]{subfig}
\usepackage{textcomp}
\usepackage{stfloats}
\usepackage{url}
\usepackage{verbatim}
\usepackage{graphicx}
\usepackage{cite}

\newcommand{\blue}[1]{{#1}}

\begin{document}

\title{Assessing Quantum Computing Performance for Energy Optimization in a Prosumer Community}
\author{Carlo Mastroianni,
Francesco Plastina,
Luigi Scarcello,
Jacopo Settino, and Andrea Vinci
\thanks{C. Mastroianni, L. Scarcello and A. Vinci are with ICAR-CNR, Via P. Bucci, 8/9 C, Rende (CS), Italy,
e-mail: \{carlo.mastroianni,luigi.scarcello,andrea.vinci\}@icar.cnr.it.}
\thanks{F. Plastina and J. Settino are with Dip. Fisica, Universit\`a della Calabria, Arcavacata di Rende (CS), Italy, and INFN - Gruppo Collegato di Cosenza, e-mail: francesco.plastina@fis.unical.it, jacopo.settino@unical.it}

}


\markboth{Transactions on Smart Grids ~Vol.~X, No.~Y, June~2023}%
{Shell \MakeLowercase{\textit{et al.}}: A Sample Article Using IEEEtran.cls for IEEE Journals}


\maketitle

\footnote{© 2023 IEEE. Personal use of this material is permitted. Permission from IEEE must be obtained for all other uses, in any current or future media, including reprinting/republishing this material for advertising or promotional purposes, creating new collective works, for resale or redistribution to servers or lists, or reuse of any copyrighted component of this work in other works.}

\begin{abstract}
The efficient management of energy communities relies on the solution of the ``prosumer problem'', 
\blue{i.e., the problem of scheduling the household loads on the basis of the user needs, the electricity prices, and the availability of local renewable energy, with the aim of reducing costs and energy waste.}
Quantum computers can offer a significant breakthrough in treating this problem thanks to the intrinsic parallel nature of quantum operations.
The most promising approach
is to devise variational hybrid algorithms, in which  quantum computation is driven by parameters that are optimized classically, in a cycle that aims at finding the best solution with a significant speed-up with respect to classical approaches. This paper provides a reformulation of the prosumer problem, allowing to address it with a hybrid quantum algorithm, namely, Quantum Approximate Optimization Algorithm (QAOA), and with a recent variant, the Recursive QAOA.
We report on an extensive set of experiments, on simulators and real quantum hardware, for different problem sizes. Results are encouraging in that Recursive QAOA is able, for problems involving up to 10 qubits, to provide optimal and admissible solutions with good probabilities, 
while the computation time is nearly independent of the system size.
\end{abstract}


\section{Introduction}
\label{secIntro}

Finding and establishing energy and climate policies are some of the greatest challenges in modern society and require, besides the transition from fossil to renewable energy sources, more efficient management of energy. The diffusion of energy communities, where prosumers can exchange energy locally, is expected to give a major contribution in this respect  \cite{nafi2016survey, giordano2020optimization}. 
Efficient energy exchange can be achieved by solving the ``prosumer problem'', which
\blue{consists in identifying optimal energy control strategies, and determining the most economical combinations for the production, purchase, and sale of energy.
More specifically, within an energy community, the objective is to schedule the prosumers' loads and storage systems in advance (e.g., one day ahead), by taking into account the user needs, the trend of the energy market, and the availability of locally produced energy. The aim is the minimization of some cost function, which can be either the monetary cost, the total energy consumption, or the consumption of brown energy. To achieve these objectives, it is crucial to maximize the energy exchanges among the community prosumers and reduce the purchase of energy from the external grid.} 


\blue{The prosumer problem is commonly expressed as a Mixed Integer Linear Programming (MILP) problem \cite{dukovska2021decentralized}, where the objective function is a linear combination of decision variables, some of which are discrete; the constraints, given by the energy balance and by the prosumers requirements, are linear as well. Several approaches have been put forward to solve such problems \cite{DR-Opt-Review}. Classical exact optimization algorithms, such as cutting plane and branch-and-bound methods~\cite{conforti2014integer}, can become infeasible for large instances, due to the NP-hard nature of the MILP problem. Metaheuristic algorithms \cite{BookDreo} (e.g., genetic algorithms, ant colony optimization, and particle swarm optimization), and machine-learning methods \cite{DR-AI-Review} can be used for large communities, when approximate solutions are acceptable.
}



In the last few years, research and industrial efforts are showing that quantum computing can offer the opportunity to approach energy optimization problems with a completely different paradigm \cite{QCForEnergySystems, Giani2021}. The main potential advantage is the computational speed-up that can be achieved by exploiting the quantum parallelism stemming from the superposition principle: a quantum state of an n-qubit register can be expressed as the superposition of $2^n$ basis states, and a quantum circuit operates in parallel over all of the basis states. Noise and decoherence issues of available quantum hardware hinder the use of pure quantum algorithms, probably for the next decades, but in the current ``noisy intermediate-scale quantum'' (NISQ) era, a valid alternative is to devise variational quantum algorithms (VQAs), in which the computation is hybrid: the most intensive computation is performed on quantum hardware, driven by a number of tunable parameters, whose values are optimized on a classical computer
\cite{biamonte2017, dunjko2016}. \blue{VQAs are seen as the quantum analogue of highly successful machine-learning methods, such as neural networks \cite{VQA}.}

\blue{In this specific application domain, one potential important advantage is that VQAs can help to tackle the uncertainty related to the production (particularly of renewable energy), consumption, and prices of energy \cite{ProsumerPriceUncertainty, StochasticUnitCommitment}. Indeed, if multiple values of input data need to be considered, included in the so-called ``uncertainty sets'', VQA algorithms can be re-executed several times, each time with different input values, starting from the optimized parameter values of previous executions. This process can be favored by the adiabatic behavior of physical processes, i.e., solutions tend to remain close for small changes in the input data.}




In this paper, following the preliminary work reported in \cite{IEEEPicom2022}, we provide a novel approach to the prosumer problem, based on the  use of one of the most renowned hybrid quantum/classical algorithms, i.e., the Quantum Approximate Optimization Algorithm (QAOA) \cite{QAOA}. The prosumer problem is first formulated as an Integer Linear Programming (ILP), where the objective is to minimize the energy cost while satisfying a number of constraints related to the maximum available energy and user requirements. The desired solution is a binary string where each bit determines whether a load should be turned on or off at a given time.
%
%
The ILP problem is then transformed into an Ising problem \cite{ising2000}, where binary variables are changed into discrete variables taking the values $\{1,-1\}$. The Ising expression defines a Hamiltonian operator, which is used to define (and, later, measure) the energy of a set of qubits, each  corresponding to one discrete variable of the Ising problem. In the latter form, the problem can be tackled by QAOA, which, using the hybrid variational technique mentioned above, aims at finding the minimum eigenvalue of the Hamiltonian operator and the corresponding eigenvector. This eigenvector represents the configuration with minimum energy -- i.e., the \textit{ground state} of the Hamiltonian -- and, at the same time, codifies the string of binary variables that determines the solution to the problem.

The evolution driving the quantum register towards the ground state is performed through a sequence of parametric quantum gates, whose parameter values are optimized by a classical optimizer. As explained in \cite{QAOA}, these gates are arranged in layers -- also called \textit{repetitions} -- and the approximated evolution aiming at the ground state is more and more precise as the number of repetitions increases. On the other hand, more repetitions require more computational time, therefore the number of repetitions is a parameter that needs to be fine-tuned.
%
The Recursive QAOA algorithm \cite{QAOA-recursive} is also often exploited to improve the quality of the solution. The objective of this variant is to reduce the size of the problem by identifying, through quantum computation, the couples of discrete variables that show maximum correlation, and eliminating one of the two. This reduction is iterated until the problem becomes tractable with a classical algorithm.



The main new contributions of our work are:
\begin{enumerate}
    \item starting from the typical formulation of the prosumer problem, we describe how it is transformed, step-by-step,
    to be given as input to QAOA, and
    how the QAOA result can be interpreted in terms of the prosumer problem; 
    \item we describe how the prosumer problem can be solved through the Recursive QAOA algorithm, which can help to improve the success probability of the experiments;
    \item we study a simple prosumer problem, 
    for which we report the results of a wide set of experiments executed on both simulator and real quantum computers provided by the IBM Quantum Experience portal\footnote{https://quantum-computing.ibm.com/}. 
\end{enumerate}

We hope that these contributions can stimulate other researchers to investigate the opportunity of adopting quantum computation to approach energy optimization problems. In this paper, however, we do not provide: (i)
 a general introduction to quantum computing basics, which can be found in excellent tutorial papers \cite{QuantumIntroNannicini, QuantumIntroRieffel}; (ii) a complete description of the QAOA and Recursive QAOA algorithms, 
 which can be found in several papers, such as \cite{QAOA} and \cite{QAOA-recursive}. 
 


At present, the size of the treatable problems is limited by the current shortcomings of quantum hardware, in terms of both size and noise.
However, we believe that the results of our experiments are promising. On the one hand, Recursive QAOA provides optimal, or at least admissible solutions, with good probabilities, in experiments with up to 10 qubits, and appears to be almost insensitive to noise. On the other hand, the experienced computation time does not increase with the number of qubits. 

%
\blue{The rest of the paper is organized as follows: Section \ref{sec:related} gives an overview of related works in this field; Section \ref{secIntroQAOA} describes how an Ising problem is transformed into the problem of finding the minimum eigenvalue of an operator, which is the goal of QAOA, and discusses the complexity of QAOA and Recursive QAOA; Section \ref{secModel} describes how the prosumer problem is transformed into an Ising problem, and therefore can be solved with the QAOA algorithm;
Section \ref{secExample} illustrates an example of a prosumer problem and reports on the form of the corresponding Hamiltonian operator, which is then given to the QAOA algorithm;
Section \ref{secResults} reports on a set of results obtained, with QAOA and Recursive QAOA,
using both simulators and the real quantum hardware, and, also based on these results, discusses the potential perspectives of quantum computation for energy scheduling problems; finally, Section \ref{secConclusions} concludes the paper.}

\section{Related~Works}
\label{sec:related}




Identifying optimal energy and climate policies are some of the greatest challenges in modern society and a transition is required from exhaustible energy sources, such as fossil and nuclear, to more sustainable modes of production and consumption. An energy transition is needed, not only in terms of energy sources, but also towards more local, decentralized and pervasive management of the energy. 
In this context, the development of prosumer communities appears to be crucial to involve citizens in the smarter exploitation of renewable sources \cite{nafi2016survey, giordano2020optimization}. 
%
Prosumers are community members that can both produce and consume the renewable energy produced locally, and exchange the surplus energy directly within the community \cite{zafar2018prosumer}.
The renewable energy surplus is traded to meet the local energy demand, thus allowing a significant reduction of energy consumption and costs. As mentioned in the introductory section, the goal of the prosumer problem is to schedule the prosumer loads in order to maximize these advantages.


\blue{In the literature, power management problems, among which unit commitment \cite{bendotti2019complexity} and hydrothermal system scheduling \cite{feng2017multi}, are often formulated as Mixed-Integer Non-Linear Programming (MINLP) problems, where some decision variables take discrete (integer or binary) values, and either the objective function or the constraints, or both, are nonlinear expressions. The prosumer problem can be expressed as a MILP problem \cite{dukovska2021decentralized}, where the objective function is linear and the constraints, given by the energy balance and by the prosumers requirements, are also linear.
Both MINLP and MILP  problems are known to be NP-hard. More specifically, MILP problems, which are the focus of this paper, are at least as complex as an Integer Linear Programming (ILP) problem that, in turn, is at least as complex as a 0--1 integer program. The last problem is convertible to the SAT optimization problem, whose time complexity is NP-hard~\cite{conforti2014integer}. As a consequence of this chain, a MILP problem is also NP-hard.}


Classical exact optimization algorithms, such as cutting plane methods and branch-and-bound~\cite{conforti2014integer}, can be used to solve MILP problems, and the literature in this specific domain is wide \cite{DR-Opt-Review}.
%
%
%
For example, in~\cite{dukovska2021decentralized}, the problem of coordinating a community of prosumers that can collaboratively share electricity is modeled as a MILP with coupling constraints. The problem is decomposed and solved using the Lagrangian duality and limited information exchange. 
A day-ahead scheduling of micro-grid resources, where the objective is to minimize the operational cost and the peak load, is presented in~\cite{teo2018near}. The proposed MILP model is solved using the CPLEX solver
in a mathematical programming language platform. 
In~\cite{novoa2019optimal}, the authors propose an optimization approach to determine the best allocation and dispatch of distributed energy resources for an energy community, while complying with electrical grid operational constraints.

In~\cite{giordano2020optimization} and~\cite{giordano2020optimizationconf}, the authors present an optimization model for energy management in a prosumer community, referred to as the Unified model. The model exploits a unique MILP problem for all the prosumers, and solves it through the Branch and Bound algorithm, which provides the optimal solution for sharing energy at the community level. 
%
In the Cascade model, discussed in~\cite{scarcello2022cascade}, the overall solution is obtained by iterating a number of MILP problems, solved sequentially.

Unfortunately, exact algorithms can become impractical for large instances~\cite{mansini2015linear, guida2015branch, krishnamoorthy2008bounds}, due to the NP-hard complexity of the problem.
To tackle the scalability issue, a Parallel approach for large communities is proposed in~\cite{scarcello2022edge}. Instead of solving a single big optimization problem, the approach divides the users into groups, and solves the optimization problems for the different groups in parallel. 
The Parallel approach helps to tackle the scalability problem, but is only able to provide a sub-optimal solution, as a prosumer belonging to a single group can coordinate and share energy only with the users of the same group.


\blue{Many heuristic algorithms can be also  employed to obtain approximate solutions, based on techniques such as genetic algorithm, tabu search, ant colony optimization, and particle swarm optimization. These algorithms perform better than exact approaches, from the computational point of view, but they do not guarantee to provide the optimal solution \cite{BookDreo}. A recent review of nature-inspired techniques for energy management systems is provided in \cite{energy-nature}, while the work in \cite{Review-DR-self} compares and evaluates a number of self-organizing algorithms for the implementation of residential demand response (DR) techniques, where the goal is to schedule the usage of devices in periods of low demand and coordinate them in order to reduce costs and avoid peaks. More recently, 
data-driven and machine-learning methods have been investigated, since they often perform better than traditional approaches. The research work in this area is also very rich. A systematic review of artificial intelligence (AI) and machine learning approaches for DR applications is provided in \cite{DR-AI-Review}, while the authors of \cite{Review-HEMS} have reviewed the AI techniques deployed for the scheduling control in home energy management systems, including neural network, fuzzy logic, and adaptive neural fuzzy inference system.}

\blue{Besides guaranteeing adequate scalability and computational performance, heuristic algorithms must also be able to address the inherent uncertainty of the actual production, consumption, and prices, which often change in real-time and deviate from the forecasts. When the uncertainty can be modeled probabilistically, it can be tackled with stochastic programming, which handles uncertain parameters in the optimization model \cite{StochasticOptimization}; otherwise, robust optimization techniques can be exploited to consider a set of worst-case scenarios and optimize the scheduling accordingly \cite{RobustOptimization}. Further heuristic approaches are based on reinforcement learning, which tries to optimize energy scheduling decisions based on past experience \cite{ReinforcementLearning}, and fuzzy logic, a mathematical framework designed to handle uncertain, imprecise or incomplete information \cite{FuzzyLogic}.
}

The quantum computing paradigm is emerging as a viable approach, thanks to its intrinsic parallelism, stemming from the laws of quantum mechanics. Specifically, the superposition and linearity principles enable exploring a search space -- the ``Hilbert space'' -- whose size is exponential with respect to the number of qubits, yet using an amount of resources, i.e., the quantum gates, which grows only polynomially, as later discussed in Section \ref{secIntroQAOA} with reference to QAOA.
The exploitation of quantum computing for energy optimization problems is a recent yet very promising research avenue. An interesting application is reported in Ref. \cite{QCForEnergySystems}, where the goal is to explore quantum approaches to specific energy problems, such as location-allocation, unit commitment, and heat exchanger.
More generally, the authors of \cite{Giani2021} discuss how quantum
computers can contribute, as the hardware
develops, to pursue renewable energy transition, by tackling problems such as the forecasting of solar and wind production, the scheduling and dispatch of renewable energy, and the reliability of power systems. In \cite{TSG-Annealing}, the authors address the optimization of power flows as a Quadratic Unconstrained Binary Optimization (QUBO) problem, using the quantum annealing paradigm, \cite{Annealing}, which is an alternative to the quantum gate paradigm adopted by major IT companies and research institutions, and was the original inspiration for the QAOA algorithm. 

Finally, in this short list of quantum-based approaches to energy problems, we mention Ref. \cite{TSG-TeachingLearning}, where the micro-grid optimal energy management problem is tackled by using a quantum variant of Teaching Learning Based Optimization, an artificial intelligence approach inspired by the learning process in a classroom, and Ref. \cite{TSG-Security}, which aims at improving the security level of smart grids, by using quantum public-key encryption and key exchange techniques that ensure mutual authentication between smart meters and the external grid.

\blue{As mentioned in the introductory section, in the current NISQ era, variational quantum algorithms (VQAs) seem to be the most promising and suitable to compensate for the noise of quantum hardware.
However, VQAs are heuristic algorithms and do not offer performance guarantees. Among them, the main advantage of QAOA is that it is known to approximate the optimal solution when the number of repetitions increases. On the other hand, the reformulation as an Ising problem, needed by QAOA, can be performed easily only for linear problems. When problems are non-linear, the theoretical background of QAOA is not exploitable, but VQAs can still be used \cite{Chang2020OnHQ}. Similarly to classical neural networks, the objective is formulated as the minimization of a cost function, achieved with an iterative approach and by the use of optimization methods, such as quantum gradient descent \cite{QuantumGradientDescent}. Today, it is impossible to perform a reliable comparison between classical approximate and quantum variational approaches, due to the heuristic nature of both, and the unavailability of results for large-size problems with quantum hardware. The advantages expected with quantum computing are related to the potential computational speed-up, and the management of uncertainty. A discussion of these aspects is provided at the end of Section \ref{secResults}, where we will also consider the complexity analysis performed in Section \ref{secIntroQAOA}, and the scalability evaluation provided in Section \ref{secScalability}.}



\section{QAOA and Recursive QAOA: goal and computational complexity}
\label{secIntroQAOA}


\blue{As already mentioned, we will leverage the QAOA algorithm for solving the prosumer optimization problem. In order for this paper to be as self-consistent as possible,  we provide, in this section, a short description of the algorithm, by focusing on how it can be used to solve a problem in the Ising form, and on its computational complexity.}

\blue{The goal of the adopted approach is to define the optimization problem in terms of a set of discrete binary variables $\{z_i\}$, taking the values $+1$ and $-1$. The solution of the problem with $N$ variables is, then, one of the $2^N$ possible strings of these values, obtained by a suitable optimization procedure.}

\blue{Each variable is associated with one of the qubits of a quantum register. In particular, $z_i$ is given by the outcome of the measurement of one of the Pauli operators, the so-called $\textbf{Z}$ observable, performed on the $i$-th qubit at the end of the algorithm. According to quantum mechanics, the measurement has, indeed, the two possible outcomes $+1$ and $-1$, which are the two eigenvalues of  $\textbf{Z}$. Correspondingly, after the measurement, the state of each qubit collapses into one of the two logical states, denoted (using Dirac notation) by $\ket 0 = [1,0]^T$ and $\ket 1= [0,1]^T$. These are the eigenstates of the $\textbf{Z}$ operator, which, as a result, can expressed, in the logical basis, as the third Pauli Matrix:}

\[
\textbf{Z} = \begin{bmatrix}
    1  & 0  \\
    0  & -1 \\
\end{bmatrix}
\]

\blue{In order to find the right $\{z_i\}$ values, the prosumer problem needs to be transformed into an Ising problem of the type:}

\begin{equation}
\label{eq:ising}
	min \ \bigg( {\sum_{i=1}^{N}{h_i \cdot z_i} - \sum_{i=1}^{N}\sum_{j=1}^{i}{J_{ij} \cdot z_i \cdot z_j}} \bigg)
\end{equation}
\blue{where $h_i$ and $J_{ij}$ are suitable real constants. This transformation is described in Section \ref{secModel}; here, we focus on a brief explanation of how the Ising problem is tackled with QAOA.}

The Ising formulation can be mapped to a diagonal operator, the Hamiltonian, built with sums and tensor products (i.e., Kronecker products) of two basic one-qubit operators, the identity \textbf{I} and the Pauli operator \textbf{Z}. For each term in (\ref{eq:ising}), the operator $\textbf{Z}_i$ substitutes the variable $z_i$, and the identity operator $\textbf{I}_i$ is added for each variable $z_i$ that does not appear in the term. Moreover, the multiplications between the $z$ variables are substituted with the tensor products between the corresponding \textbf{Z} operators. For example, with $N$=$4$ the term $z_2$$\cdot$$z_3$ is substituted with $\textbf{I}_1 \otimes \textbf{Z}_2 \otimes \textbf{Z}_3 \otimes \textbf{I}_4$ or, more succinctly, $\textbf{I}_1 \textbf{Z}_2 \textbf{Z}_3 \textbf{I}_4$ or, even more briefly, $\textbf{Z}_2 \textbf{Z}_3$, where the identity operators are implicit.  With these rules, the Hamiltonian operator that corresponds to expression (\ref{eq:ising}) is:

\begin{equation}
\label{eq:IsingHamiltonian}
	\textbf{H} = {\sum_{i=1}^{N}{h_i \cdot \textbf{Z}_i} - \sum_{i=1}^{N}\sum_{j=1}^{i}{J_{ij} \cdot \textbf{Z}_i \otimes \textbf{Z}_j}}
\end{equation}

Now, the problem is to find the minimum eigenvalue(s) of the operator (\ref{eq:IsingHamiltonian}), which corresponds to finding the string of values of $z$ variables that minimizes the Ising expression (\ref{eq:ising}).
%
%
%
Indeed, the Hamiltonian operator, thanks to the properties of the tensor product\footnote{The tensor product of two diagonal matrices is still diagonal.}, is always represented as a diagonal matrix, and the eigenstate corresponding to the minimum eigenvalue, i.e., the ground state, 
gives the solution to the problem. \blue{For example, let us assume that our Ising problem is:}

\begin{equation*}
\label{eq:ising-ex}
	min \ \big(z_1 + 2 z_3 - 4 z_1 z_2 -2 z_2 z_3 \big)
\end{equation*}

\blue{and, then, that the Hamiltonian of the problem is:}

\begin{multline}
\label{eq:Hamiltonian-ex}
	\textbf{H} = 
 (\textbf{Z}_1 \otimes \textbf{I}_2 \otimes \textbf{I}_3) 
 +2\ (\textbf{I}_1 \otimes \textbf{I}_2 \otimes \textbf{Z}_3)  \\
 -4\ (\textbf{Z}_1 \otimes \textbf{Z}_2 \otimes \textbf{I}_3) 
 -2\ (\textbf{I}_1 \otimes \textbf{Z}_2 \otimes \textbf{Z}_3) 
\end{multline}
\blue{The matrix representation of \textbf{H} is:}
\[
\textbf{H} = \begin{bmatrix}
    -3 & 0 & 0 & 0 & 0 & 0 & 0 & 0 \\
    0 & -3 & 0 & 0 & 0 & 0 & 0 & 0 \\
    0 & 0 & 9 & 0 & 0 & 0 & 0 & 0 \\
    0 & 0 & 0 & 1 & 0 & 0 & 0 & 0 \\
    0 & 0 & 0 & 0 & 3 & 0 & 0 & 0 \\
    0 & 0 & 0 & 0 & 0 & 3 & 0 & 0 \\
    0 & 0 & 0 & 0 & 0 & 0 & -1 & 0 \\
    0 & 0 & 0 & 0 & 0 & 0 & 0 & -9
\end{bmatrix}
\]

\blue{The Hamiltonian shows $2^N$  eigenvalues, i.e., the diagonal elements of the matrix, where $N$ is the number of qubits.
Each basis state is an eigenvector that corresponds to a possible assignment of the discrete variables $\{z_i\}$, associated with the qubits, and the corresponding eigenvalue gives the value of the cost function. In this case, the minimum eigenvalue is $-9$, obtained for the eigenstate $[0,0,0,0,0,0,0,1]^T$, which, in turn, corresponds to the basis state $\ket{111} = \ket{1} \otimes \ket{1} \otimes \ket{1}$ = $[0,1]^T \otimes [0,1]^T \otimes [0,1]^T$ = $[0,0,0,0,0,0,0,1]^T$, and, thus, to the solution $z_1=-1$, $z_2=-1$, and $z_3=-1$.}

\begin{figure}[!tb]
	\centering
	\includegraphics[width=0.95\columnwidth, trim=0cm 0cm 0cm 0cm]{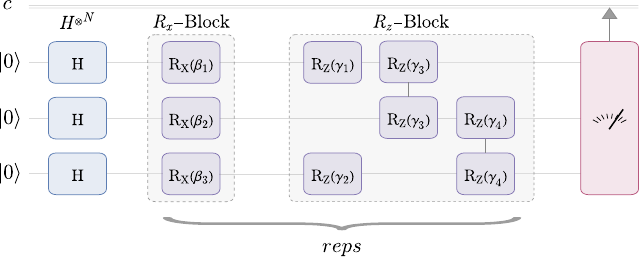}
	\caption{\blue{Example of a QAOA circuit. The objective is to tune the values of the parameters $\beta$ and $\gamma$ and prepare, before measurement, the eigenstate that corresponds to the minimum eigenvalue of the Hamiltonian defined in (\ref{eq:Hamiltonian-ex})}.}
	\label{fig:QAOAcomplexity}
\end{figure}

\blue{The objective of a QAOA is to find the eigenstate corresponding to the minimum eigenvalue of the Hamiltonian operator, without computing all the $2^N$ diagonal elements.
This is accomplished through a hybrid iterative algorithm in which a quantum circuit prepares an output state  through a set of parametric gates, and a classical routine is used to optimize the parameter values until the prepared state corresponds to the ground state of the Hamiltonian.}

\blue{A full description of the details of QAOA cannot be given here for reasons of space. Instead, we focus on the computational complexity of the algorithm, in terms of the number of gates and the depth of the circuit. The QAOA quantum circuit for the problem expressed in Eq. (\ref{eq:Hamiltonian-ex}) is depicted in Figure (\ref{fig:QAOAcomplexity}), which shows that there are three main groups of quantum gates, where the last two blocks are repeated a number of times referred to as \textit{reps}.}
%
%
%
%
\blue{The number of gates can be computed for the different blocks of the circuit: the first block requires $N$ Hadamard gates; the second block (the $R_x$ block) requires X-rotation gates, one for each qubit: the third block (the $R_z$ block) requires at most $N$ single qubit Z-rotation gates (one for each term $\textbf{Z}_i$ of the Hamiltonian) and at most $N \cdot N /2$ two-qubit gates (one for each term $\textbf{Z}_i \textbf{Z}_j$ of the Hamiltonian), i.e., $\mathcal{O} (N^2)$ two-qubit gates. 
The second and third stages are repeated a number of times equal to \textit{reps}, and at the end of the last repetition, a measurement is performed on the entire qubit register.}

\blue{The QAOA circuit is executed a number of times equal to \textit{shots} in order to obtain a statistically significant distribution of measurement results. Overall, the number of needed gate executions for a single iteration of the hybrid algorithm is $\mathcal{O} (shots \cdot reps \cdot N^2)$, i.e., still $\mathcal{O} (N^2)$. Since the gates can be executed in parallel on different qubits, the depth of the circuit is $\mathcal{O} (N)$. This is an upper limit, obtained when the Ising expression contains all the possible couples of discrete variables $z$. In real cases, where only a subset of those couples are present, the depth grows more slowly, or is nearly constant.}

\blue{The hybrid algorithm requires a number of iterations to converge, which is very difficult to estimate. The hope is that the number of iterations increases at most polynomially with respect to $N$. This hope stems from the observation that the quantum gates operate on a Hilbert space whose number of dimensions is exponential with respect to the number of qubits. However, at present, there is no proof regarding the expected number of iterations, neither for QAOA nor for any other variational hybrid algorithm, which justifies the large amount of research work that is devoted to the experimental analysis of these algorithms.}


Besides using the original QAOA algorithm, we also tested the performance of a recent variant, namely, the Recursive QAOA algorithm \cite{QAOA-recursive}. The main idea is to exploit QAOA to find the possible correlations between pairs of discrete variables, and reduce the size of the problem down to a number of variables, denoted as $num\_min\_var$, after which the problem is given to a classical optimizer that finds the best solution.
After the first execution of QAOA, the expected value of each term $Z_i \otimes Z_j$ in Eq. (\ref{eq:IsingHamiltonian}) is measured, and the term $Z_{\hat{i}}\otimes Z_{\hat{j}}$ with the largest absolute value, which corresponds to the maximum correlation or anti-correlation, is identified\footnote{If the value is close to 1, it means that the two discrete variables often assume the same value in a solution (both 1 or both -1), while if the value is close to -1, it means that the variables often assume opposite values.}. This gives the possibility of eliminating one of the two variables, for example, $Z_{\hat{j}}$, by substituting it with $Z_{\hat{i}}$ or $-Z_{\hat{i}}$ in (\ref{eq:IsingHamiltonian}), depending on the two variables being correlated or anti-correlated. Then, QAOA is executed again with the new Hamiltonian operator, and the procedure is iterated until the number of variables is reduced to $num_min_var$. At this point, the problem is solved classically, and the optimal solution hopefully is the best solution to the original problem or at least a sub-optimal admissible solution.

The efficacy of Recursive QAOA is good if strong correlations exist between pairs of variables. Since at every iteration the strength of correlation decreases, as better correlations were found at the previous steps, it is reasonable to expect that more iterations can decrease the quality of the solution found with the final classical optimization step. On the other hand, classical optimization is not scalable; therefore, when the number of qubits is large, the number of iterations should be large enough to reduce the size of the problem until it can be effectively tackled classically. The number of iterations must be set by balancing these two contrasting considerations.

\blue{Since Recursive QAOA repeats the execution of QAOA a number of times equal to $N - num\_min\_var$, the complexity is $
\sum_{i=0}^{N-num\_min\_var} \mathcal{O}(N-i)^2=\mathcal{O}(N^3)$, achieving polynomial time complexity. It is worth noting that the same considerations made about the number of iterations needed by QAOA to converge still hold for the Recursive QAOA.}



\section{Formulating the Prosumer Problem for QAOA}
\label{secModel}

%
%

As anticipated in the introductory section, this paper describes a general approach to transform a prosumer problem into a problem that can be solved with the QAOA algorithm.
The prosumer problem is defined as the problem of minimizing the energy cost incurred by the users of an energy community, while satisfying a set of constraints, i.e., addressing the user energy requirements and providing a feasible scheduling for a set of 
schedulable interruptible loads.
A schedulable load is an electric load that can be scheduled within a time interval, in accordance with the user preferences; moreover, it is interruptible when it can be stopped and resumed at a different time. \blue{In the following, we define a simple version of the prosumer problem, involving a number of prosumers of an energy community. The aim is to showcase the steps needed for the problem reformulation in terms of a Hamiltonian operator. At the end of the section, we will mention the further specifications and constraints that can be added to the prosumer problem, and how they can be tackled to exploit the QAOA algorithm.}


Some necessary definitions are provided in the following:
\begin{labeling}{xxxxx}
    \item [$S_U$] set of users belonging to the community;
    \blue{
    \item [$S_{L_u}$] set of schedulable loads of user $u \in S_U$;
	\item  [$L_u$] number of schedulable loads of user $u \in S_U$;
    }
    \item [$S_H$] set of scheduling hours;
    \item [$H$] number of scheduling hours;
    \blue{
    \item [$\delta_{l_u}$] working time of the load $l \in S_{L_u}$, i.e., number of hours for which the load must be switched on;
	\item [$E_{l_u}$] power consumption of the schedulable load $l \in S_{L_u}$ for a user $u \in S_U$;
	\item [$E_{max,u}$] maximum nominal power available to user $u$;
	\item [$p^h$] cost of the electrical energy imported by the community (at a discounted tariff) from the grid at the hour $h \in S_H$;
	\item [$x_{l_u}^h$] state of the schedulable load $l \in S_{L_u}$ at hour $h \in S_H$ ($1$ = on; $0$ = off).}
\end{labeling}



\blue{The problem is to minimize the following objective function:}


\blue{
\begin{equation}
\label{eq:objFunc}
	\ C = {\sum_{h \in S_H}
 {p^h \cdot \Big(
 \sum_{u \in {S_U}}\sum_{l_u\in {S_{L_u}}}E_{l_u} \cdot {x_{l_u}^{h}}
\Big)}}
\end{equation}
}


\blue{Eq. (\ref{eq:objFunc}) defines the daily energy cost afforded by the energy community, which is minimized by scheduling the states (on or off) of the loads $x_{l_u}^h$. The cost is computed by multiplying the electrical tariffs $p^h$, granted to the community at the different hours $h$, by the power consumptions $E_{l_u}$ of active loads. The following constraints need to be satisfied:}


\blue{
\begin{align}
\label{eq:constraintEmax}
	&\sum_{l_u\in S_{L_u}}({x_{l_u}^h} \cdot E_{l_u}) \leq E_{max,u}
    ~~~ \forall h\in S_H, \forall u\in S_U\\
\label{eq:constraintDelta}
	&\sum_{h\in S_H}{x_{l_u}^h}=\delta_{l_u} ~~~ \forall u \in S_U, \forall l_u \in {S_{L_u}}
\end{align}
}

\blue{Inequalities (\ref{eq:constraintEmax}) force the value of the energy supplied to the loads, for each user and at each hour, to be lower or equal than the nominal power available at the user's electrical system.
Equations (\ref{eq:constraintDelta}) ensure that each load $l$ of each user $u$ is switched on exactly for $\delta_l$ hours.} 

\blue{For this optimization problem, the number of binary variables $x_{l_u}^h$ is equal to $H \cdot \sum_{u \in S_{U}}(L_{u})$. The values of energy, $E_{max,u}$ and $E_{l_u}$, are expressed as integers. This is not a limitation because, if fractional values are needed, they can be converted to integers through multiplication by an appropriate factor.}

\blue{As discussed in Section \ref{secIntroQAOA}, the problem needs to be transformed into an Ising problem, of the type shown in Expression (\ref{eq:ising}). This transformation is described in the following.
The inequalities (\ref{eq:constraintEmax}) are converted into equations,
by adding extra integer variables $E_{res,u}^h$ that represent, for each user and at each hour, the residual energy that is not used, with $0 \le E_{res,u}^h \le E_{max,u}$:}


\blue{
\begin{equation}
\label{eq:conversioneInequToEqu}
    \sum_{l_u\in S_{L_u}}({x_{l_u}^h} \cdot E_{l_u}) + E_{res,u}^h = E_{max,u}  ~~~ \forall h\in S_H, \forall u\in S_U\\
\end{equation}
}

\blue{The integer values of $E_{res,u}^h$ can be expressed with the use of a number of slack binary variables equal to $M_{u}$, with $M_u = \lceil log(N_{res,u}) \rceil$, where $N_{res,u}$ is the number of integer values that $E_{res,u}^h$ can take, i.e., 
$N_{res,u} = E_{max,u} + 1$. By using the slack variables, denoted with $y$, $E_{res,u}^h$ can be expressed as:}

\blue{
\begin{multline}
\label{eq:conversioneIntToBin}
    E_{res,u}^h = \sum_{m_u = 1}^{M_u-1}(2^{m_u-1} \cdot y_{m_u}^h) + (N_{res,u}-2^{M_u-1}) \cdot y^h_{M_u}\\ ~~~ \forall h\in S_H, \forall u\in S_U
\end{multline}
}

\blue{
and the inequalities (\ref{eq:constraintEmax}) can be converted into the equations:
}

\blue{
\begin{multline}
    \sum_{l_u\in S_{L,U}}({x_{l_u}^h} \cdot E_{l_u})  \\ + \sum_{m_u = 1}^{M_u-1}(2^{m_u-1} \cdot y_{m_u}^h) + (N_{res,u}-2^{M_u-1}) \cdot y^h_{M_u} = E_{max,u} ~~~ \\ \forall h\in S_H, \forall u\in S_U
\end{multline}
}

\blue{In this way, we added  $H \cdot \sum_{u \in S_{U}}(M_{u})$ binary slack variables, and the total number $N$ of binary variables (including the original variables $x$) becomes:}

\blue{
\begin{equation}
N = H \cdot \Big( \sum_{u \in S_{U}}(L_{u}) + \sum_{u \in S_{U}}(M_{u}) \Big)
\end{equation}
}



The next step is to build a function that incorporates the original objective function of Eq. (\ref{eq:objFunc}) and a sum of penalties, each corresponding to a constraint. 
The addition of these penalties creates an augmented objective function to be minimized.
If the penalty terms can be driven to zero, through a proper setting of the binary variables, the augmented objective function becomes the original function to be minimized.
That is, the penalties equal zero for feasible solutions and equal some positive penalty amount for infeasible solutions \cite{glover2022quantum}.  

The prosumer problem now becomes the minimization of a Quadratic Unconstrained Binary Optimization (QUBO) expression\footnote{This expression includes constant terms, which should not be present in a QUBO expression. However, they can be ignored because they do not play a role in the minimization.}:

\blue{
\begin{multline}
\label{eq:proQUBO}
	min \bigg(
 {\sum_{h \in S_H} {p^h \cdot \Big(\sum_{u \in {S_U}}\sum_{l_u\in {S_{L_u}}}E_{l_u} \cdot {x_{l_u}^{h}} \Big)}}\\
	+A \cdot \bigg\{
 \sum_{h\in S_H} \sum_{u\in S_U}
	\bigg[
  \sum_{l_u\in S_{L,U}}({x_{l_u}^h} \cdot E_{l_u}) + \sum_{m_u = 1}^{M_u-1}(2^{m_u-1} \cdot y_{m_u}^h) \\+ (N_{res,u}-2^{M_u-1}) \cdot y^h_{M_u} - E_{max,u}
 \bigg]^2\\
	+ \sum_{u\in S_U}\sum_{l_u\in S_{L_u}}\bigg[
 \sum_{h\in S_H}{x_{l_u}^h} - \delta_{l_u}
 \bigg]^2\bigg\}\bigg)
\end{multline}
}


In Eq. (\ref{eq:proQUBO}), $A$ is a penalty coefficient computed as in Eq. (\ref{eq:costanteA}), where $C_{up}$ and $C_{low}$ are, respectively, the upper and lower bounds of the cost function $C$, defined in Eqs. (\ref{eq:costanteAUP}) and (\ref{eq:costanteALOW}). 

\blue{
\begin{align}
    \label{eq:costanteA}
    &A = 1.0 + (C_{up} - C_{low})\\
    \label{eq:costanteAUP}
    &C_{up} = 
    {\sum_{h \in S_H}
 {p^h \cdot \Big(
 \sum_{u \in {S_U}}\sum_{l_u\in {S_{L_u}}}E_{l_u} \cdot {1}
\Big)}}\\
    \label{eq:costanteALOW}
    &C_{low} = {\sum_{h \in S_H}
 {p^h \cdot \Big(
 \sum_{u \in {S_U}}\sum_{l_u\in {S_{L_u}}}E_{l_u} \cdot {0}
\Big)}}
\end{align}
}


The value of the non-negative constant A is sufficiently large to ensure that the minimum of the objective function is obtained when all the constraints are matched. This occurs because a possible penalty, deriving from the non-satisfaction of a constraint, cannot be compensated by the decreasing of the original objective function that is allowed by the missed constraint satisfaction.


\blue{Now we can transform the QUBO problem expressed in Eq. (\ref{eq:proQUBO}) into an Ising problem, by substituting each binary variable with a discrete variable, as follows:}

\blue{
\begin{equation}
\label{eq:substitution}    
x_{l_u}^h =\frac{1-z_{l_u}^h}{2}, ~y_{m_u}^h =\frac{1-z_{m_u}^h}{2}
\end{equation}
}
%
%
\blue{
At this point, an Ising problem as in Eq. (\ref{eq:ising}) has been obtained: the values $\{0,1\}$ of a binary variable correspond to the values $\{+1,-1\}$ of the matching discrete variable\footnote{Note that ${(x_{l_u}^h)}^2$=$x_{l_u}^h$=$\frac{1-z_{l_u}^h}{2}$, since the square of a binary variable equals the binary variable itself, and the same occurs with ${(y_{m_u}^h)}^2$. This is the reason why there are no terms ${(z_i)}^2$ in Eq. (\ref{eq:ising}).}.}

%

%

\blue{As discussed in Section \ref{secIntroQAOA}, the problem becomes that of finding the minimum eigenvalue of the Hamiltonian operator defined in (\ref{eq:IsingHamiltonian}). A register of $N$ qubits is prepared by QAOA to achieve with maximum probability the ground state, in which each qubit, after measurement, collapses into one of the basis states $\ket{0}$ or $\ket{1}$. The solution to the problem is obtained by setting each binary variable
to $0$ (if the corresponding qubit collapses to $\ket{0}$, the measurement result being the eigenvalue +1) or $1$ (if the corresponding qubit collapses to $\ket{1}$, the measurement result being $-1$).}

In general, the output state prepared by the quantum circuit is a superposition of various basis states. The objective, then, is not necessarily to prepare the ground state, but a state in which the amplitudes of the basis states that represent the solution are larger than the other amplitudes, so that a measurement provides the desired solution with a large probability.

\blue{As mentioned above, the formulation of the prosumer problem can be more complex and include more constraints, and more terms in the objective function, to take into account, e.g.: (i) the renewable energy produced by the prosumers, which can be shared within the community; (ii) the presence of different prices for buying/selling energy from/to the grid and within the community; (iii) the presence of energy storage systems. More complete formulations can be found in the literature \cite{belli2017unified, cui2018two}. These further aspects are not described here for the sake of readability; however, they can be tackled in the same way as shown before: inequalities must be transformed into equations, each integer variable must be converted into a number of binary variables, the constraints must be included in the objective function, and variable substitutions must be performed to achieve an Ising problem formulation.}

\section{Example of a Specific Prosumer Problem}
\label{secExample}

This section provides and illustrates an example of a prosumer problem, which will be used for the experiments. The example is related to one user of an energy community, equipped with two schedulable loads, i.e., $L=2$, while the number of hours, $H$, ranges between two and five.
The aim is to perform experiments
in order to assess the ability of the QAOA algorithm to solve the problem both in a simulation environment and with real quantum hardware.
The constants are set as follows:
\begin{itemize}
    \item Working times of the loads:

	     $\delta_1$ = 2 [h];\\
	   $\delta_2$ = 1 [h];
 
     \item Power consumption of the loads and nominal power of the system:
	    $E_1$ = 1 [kW];\\
	    $E_2$ = 2 [kW];\\
	    $E_{max}$ = 3 [kW];
    \item Hourly cost of the energy:
	    $p^1$ = 21 [\euro \ cent / kWh];\\
	    $p^2$ = 21 [\euro \ cent / kWh];\\
	    $p^3$ = 22 [\euro \ cent / kWh];\\
	    $p^4$ = 23 [\euro \ cent / kWh];\\
	    $p^5$ = 24 [\euro \ cent / kWh];
\end{itemize}

The function to minimize, whose general expression is given in Eq. (\ref{eq:objFunc}), here becomes:
\begin{equation}
\label{eq:objFuncExample}
		\ C =  E_{1} \cdot \sum_{h=1}^H(p^h \cdot x_1^h) + 
		E_{2} \cdot \sum_{h=1}^H(p^h \cdot x_2^h)
\end{equation}

The constraints expressed in Eqs. (\ref{eq:constraintEmax}) and (\ref{eq:constraintDelta}) become:
\begin{align}
\label{eq:emax}
	&{x_1^h} \cdot E_{1} + {x_2^h} \cdot E_{2} \leq E_{max}, \ h=1~...~H \\
	&\sum_{h=1}^H(x_1^h) = \delta_1, ~\sum_{h=1}^H(x_2^h) = \delta_2
\end{align}


As in Eq. (\ref{eq:conversioneInequToEqu}), the inequality constraints are converted into equations:
\begin{equation}
\label{eq:Eres}
	{x_1^h} \cdot E_{1} + {x_2^h} \cdot E_{2} + E_{res}^{h} = E_{max}, \ h=1~...~H
\end{equation}


At this point, we convert the integer variables $E_{res}^h$ into binary variables. Since the integer variables can take only values in the range $[0; E_{max}]$, the parameters $N_{res,u}$ and $M_u$ of Eq. (\ref{eq:conversioneIntToBin}) are equal to 4 and 2, respectively. By substituting these values in Eq. (\ref{eq:conversioneIntToBin}), we obtain:
\begin{equation}
\label{eq:Eres2}
    E_{res}^{h} = y_1^h + 2 \cdot y_2^h, \ h=1~...~H
\end{equation}

The constraint (\ref{eq:emax}) can now be expressed in terms of the original load variables $x$ and the slack variables $y$, as follows:
\begin{align}
    &{x_1^h} \cdot E_{1} + {x_2^h} \cdot E_{2} + y_1^h + 2 \cdot y_2^h = E_{max}, \ h=1~...~H
\end{align}



The number of binary variables is equal to $L \cdot H + M \cdot H$, where the first term is the number of the original variables and the second term is the number of slack variables. In this problem, we introduced a simplification, in order to make the problem solvable with a reduced number of qubits and then executable with currently available quantum computers; namely, the maximum amount of power required by the loads never exceeds the rated power $E_{max}$. This
eliminates the necessity of slack variables, thus reducing the number of binary variables --- and qubits -- to $L \cdot H$.

The problem is transformed into a QUBO problem, with the procedure discussed in Section \ref{secModel}. As an example, when $H = 4$, the function to minimize becomes:
\begin{multline}
\label{eq:QUBOproblem}
    (
    21 \cdot x_1^1 
    + 21 \cdot x_1^2 
    + 22 \cdot x_1^3 
    + 23 \cdot x_1^4) \cdot 1\\
    +(21 \cdot x_2^1 
    + 21 \cdot x_2^2 
    + 22 \cdot x_2^3 
    + 23 \cdot x_2^4) \cdot 2\\
    +A \cdot \{
    [{x_1^1 + x_1^2 + x_1^3 + x_1^4} - 2]^2\\
    +[{x_2^1 + x_2^2 + x_2^3 + x_2^4} - 1]^2
    \}
\end{multline}%

where the penalty coefficient $A$ is computed using Eq. (\ref{eq:costanteA}): 

\begin{multline*}
\label{eq:costanteAEsempio}
    A = 1.0 + \{21 \cdot 1 + 21 \cdot 1 + 22 \cdot 1 + 23 \cdot 1 + 42 \cdot 1 + 42 \cdot 1 + 44 \cdot 1 + 46 \cdot 1\}\\
    - \{21 \cdot 0 + 21 \cdot 0 + 22 \cdot 0 + 23 \cdot 0 + 42 \cdot 0 + 42 \cdot 0 + 44 \cdot 0 + 46 \cdot 0\} = 262
\end{multline*}
%



As discussed before, the Hamiltonian operator \textbf{H} is obtained with the substitution $x$ (or $y$) = $(1-z)/2$, and by putting $Z$ operators in place of the $z$ discrete variables:

\begin{multline}
    \textbf{H} =  -283 \cdot Z_1
    - 283 \cdot Z_2
    - 284 \cdot Z_3
    - 285 \cdot Z_4
    - 10.5 \cdot Z_5\\
    - 10.5 \cdot Z_6
    - 11 \cdot Z_7
    - 11.5 \cdot Z_8
    + 131 \cdot Z_1Z_2
    + 131 \cdot Z_1Z_3\\
    + 131 \cdot Z_2Z_3
    + 131 \cdot Z_1Z_4
    + 131 \cdot Z_2Z_4
    + 131 \cdot Z_3Z_4
    + 131 \cdot Z_5Z_6\\
    + 131 \cdot Z_5Z_7
    + 131 \cdot Z_6Z_7
    + 131 \cdot Z_5Z_8
    + 131 \cdot Z_6Z_8
    + 131 \cdot Z_7Z_8
\end{multline}



The QAOA algorithm can now be executed to obtain the ground state of \textbf{H}, which encodes the values of the binary variables that minimize the QUBO function (\ref{eq:QUBOproblem}).

The problem with $H = 4$ has 24 admissible solutions. For ten of these solutions, Table \ref{tab:probability2} reports the values of the eight binary variables, and the corresponding values of the cost function, obtained from Eq. (\ref{eq:objFuncExample}). The best solutions are the first two in the table: they are obtained when the first load, with a power consumption of $2\ kW$, is activated either in the first or in the second hour, and the second load, with a power consumption of $1\ kW$, is scheduled in both the first and the second hour. The corresponding global energy cost is 0.84 \euro.


\begin{table}[!tb]
 \caption{Some admissible solutions and values of the cost function for the example prosumer problem}
\label{tab:probability2}
\centering

\begin{tabular}{|c|c|c|c|c|c|c|c|c|c|}
\hline
\begin{tabular}[c]{@{}c@{}}admissible\\ solution\end{tabular} & $x_1^1$ & $x_1^2$ & $x_1^3$ & $x_1^4$ & $x_2^1$ & $x_2^2$ & $x_2^3$ & $x_2^4$ & \begin{tabular}[c]{@{}c@{}}cost\\ {[}€ cent{]}\end{tabular} \\ \hline
1                                                           & 1                       & 0                       & 0                       & 0                       & 1                       & 1                       & 0                       & 0                       & {\color[HTML]{333333} 84}                                                          \\ \hline
2                                                           & 0                       & 1                       & 0                       & 0                       & 1                       & 1                       & 0                       & 0                       & {\color[HTML]{333333} 84}                                                          \\ \hline
3                                                           & 1                       & 0                       & 0                       & 0                       & 1                       & 0                       & 1                       & 0                       & {\color[HTML]{333333} 85}                                                          \\ \hline
4                                                           & 0                       & 1                       & 0                       & 0                       & 1                       & 0                       & 1                       & 0                       & {\color[HTML]{333333} 85}                                                          \\ \hline
5                                                           & 1                       & 0                       & 0                       & 0                       & 0                       & 1                       & 1                       & 0                       & {\color[HTML]{333333} 85}                                                          \\ \hline
6                                                           & 0                       & 1                       & 0                       & 0                       & 0                       & 1                       & 1                       & 0                       & {\color[HTML]{333333} 85}                                                          \\ \hline
7                                                           & 0                       & 0                       & 1                       & 0                       & 1                       & 1                       & 0                       & 0                       & {\color[HTML]{333333} 86}                                                          \\ \hline
8                                                           & 1                       & 0                       & 0                       & 0                       & 1                       & 0                       & 0                       & 1                       & {\color[HTML]{333333} 86}                                                          \\ \hline
9                                                          & 0                       & 1                       & 0                       & 0                       & 1                       & 0                       & 0                       & 1                       & {\color[HTML]{333333} 86}                                                          \\ \hline
10                                                          & 1                       & 0                       & 0                       & 0                       & 0                       & 1                       & 0                       & 1                       & {\color[HTML]{333333} 86} \\ \hline
\end{tabular}
\end{table}


\section{Results}
\label{secResults}

In this section, we present a set of experimental results obtained through the resolution of the prosumer problem with QAOA. The objectives of the experiments are the following:

\begin{itemize}
    \item assess the performance of the algorithm in a simulation environment. To this aim, we executed the algorithm using the IBM Qasm simulator;
    \item investigate the impact of noise on the performance. We performed a set of experiments in which the Qasm simulator was configured using the noise model and the qubit coupling of real quantum hardware, specifically, \textit{ibmq\_montreal}, an IBM quantum device of type Falcon R4, equipped with 27 qubits;
    \item test the real quantum hardware, using the \textit{ibmq\_montreal} device mentioned in the previous point, and compare its performances to those obtained with simulation;
    \item assess the performance of the Recursive QAOA algorithm;
    \item analyze the scalability behavior of the quantum computing approach and compare it to a classical approach.
\end{itemize}

The performances were analyzed using the prosumer problem described in Section \ref{secExample} when varying: (i) the size of the problem, i.e., the number of qubits, which in our case is equal to twice the number of hours at which the loads can be scheduled; (ii) the number of repetitions of the QAOA algorithm: as mentioned in the introductory section, the increase in the number of repetitions leads to a more precise solution but requires more computational time.
The simulation results were averaged over 20 runs, while the experiments on real hardware were executed once, due to the long waiting times experienced on real devices, as these devices are made available to numerous users and institutions worldwide.
    
We evaluated the following performance indices:
\begin{itemize}
    \item the \textit{success probability}, $P_{best}$, computed as the frequency at which the final measurement produced the optimal solution of the problem. The number of measurements (\textit{shots}) was set to 4096 for all the experiments. With the Recursive QAOA algorithm, the final step is classical, and only one solution is given as an output: therefore this index is computed as the fraction of runs for which the obtained solution is equal to the optimal solution;
    \item the \textit{probability of admissible solution}, $P_{adm}$, defined as the probability that the final measurement gives an admissible solution, i.e., a solution (optimal or non-optimal) that satisfies the constraints. With Recursive QAOA, this index is defined as the fraction of runs for which the obtained solution is admissible;
    \item the \textit{computing time} of the algorithm execution.
\end{itemize}

The rest of this section discusses the most interesting results of the experiments.

\subsection{Performance of QAOA on Simulator and Real Hardware}
\label{secSimReal}

\begin{figure}[!tb]
	\centering
	\includegraphics[width=0.95\columnwidth, trim=0cm 0.5cm 0cm 0cm]{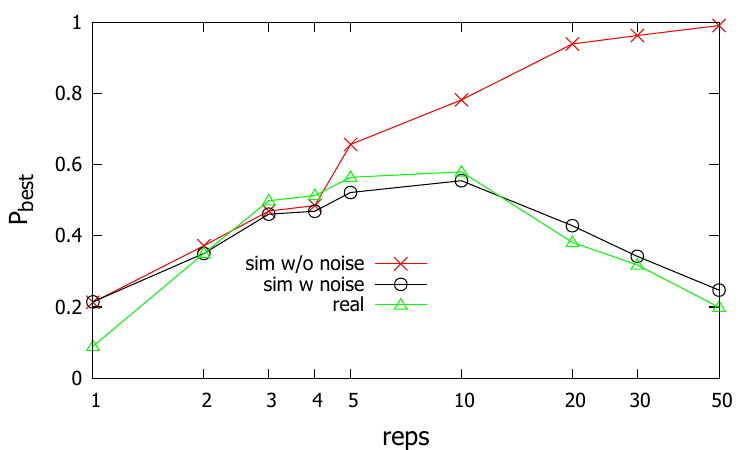}
	\caption{Probability $P_{best}$ with 4 qubits. It coincides with the probability $P_{adm}$, since in this case all feasible solutions are also optimal solutions.}
	\label{fig:confronto_4qubit_qaoa_probfvalEsatta}
\end{figure}

The first objective was to verify whether QAOA is able to achieve the optimal solution or at least a sub-optimal but admissible solution. Figure \ref{fig:confronto_4qubit_qaoa_probfvalEsatta} reports the success probability, $P_{best}$, for the 4-qubit problem, using noiseless and noisy simulation and the real quantum hardware. In this and in the following figures, the results are plotted versus the number of repetitions of the QAOA circuit, denoted as ``reps''. In the 4-qubit case, there are only two admissible solutions, which are also optimal, therefore this index coincides with the index $P_{adm}$.
We can observe that:
\begin{enumerate}
    \item using a noiseless simulator, the success probability increases with the number of repetitions and approaches 1.0 with 20 or more repetitions. This confirms the theoretical prediction \cite{QAOA};
    \item when using a noisy simulation, the effect of noise is amplified by the number of repetitions. Therefore, there is an optimal value of \textit{reps}, which appears to be around 10, after which the success probability begins to decrease. This suggests that, depending on the size of the problem and the impact of noise, the value of repetitions must be tuned carefully;
    \item the performances on the real hardware are compatible with those of the noisy simulation, despite some differences, probably due to the fact that only one test was performed on the quantum device, while simulation results are averaged over 20 runs. 
\end{enumerate}

The last observation suggests that the noisy simulation is a good approximation of the results expected on real hardware. In the following, we will report more results obtained on a real device, but also many results obtained on a noisy simulator, which are statistically more significant since they are averaged over many experiments.

\begin{figure}[!tb]
	\centering
	\includegraphics[width=0.95\columnwidth, trim=0cm 0.5cm 0cm 0cm]{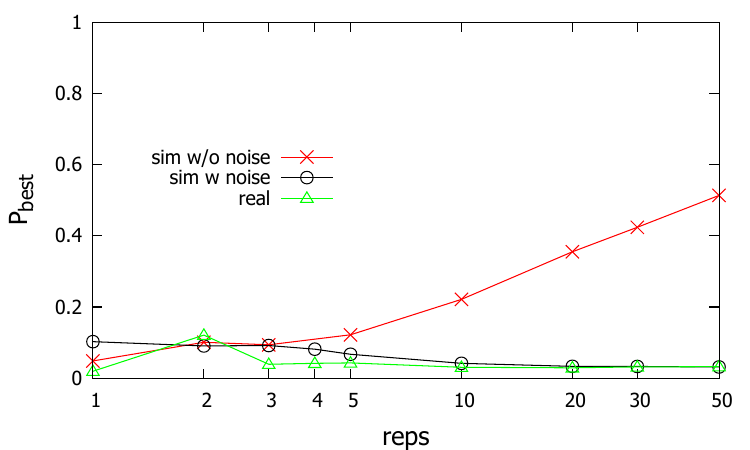}
	\caption{Probability $P_{best}$ with 6 qubits.}
	\label{fig:confronto_6qubit_qaoa_probfvalEsatta}
\end{figure}

\begin{figure}[!tb]
	\centering
	\includegraphics[width=0.95\columnwidth, trim=0cm 0.5cm 0cm 0cm]{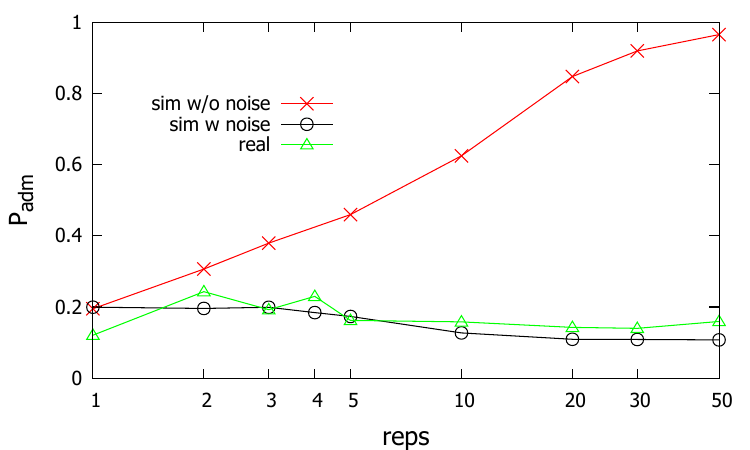}
	\caption{Probability $P_{adm}$ with 6 qubits.}
	\label{fig:confronto_6qubit_qaoa_probSuccess}
\end{figure}

\begin{figure}[!tb]
	\centering
	\includegraphics[width=0.95\columnwidth, trim=0cm 0.5cm 0cm 0cm]{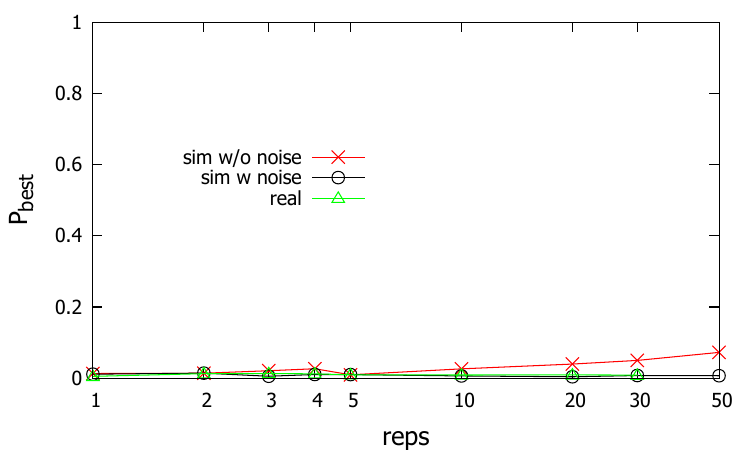}
	\caption{Probability $P_{best}$ with 8 qubits.}
	\label{fig:confronto_8qubit_qaoa_probfvalEsatta}
\end{figure}

\begin{figure}[!tb]
	\centering
	\includegraphics[width=0.95\columnwidth, trim=0cm 0.5cm 0cm 0cm]{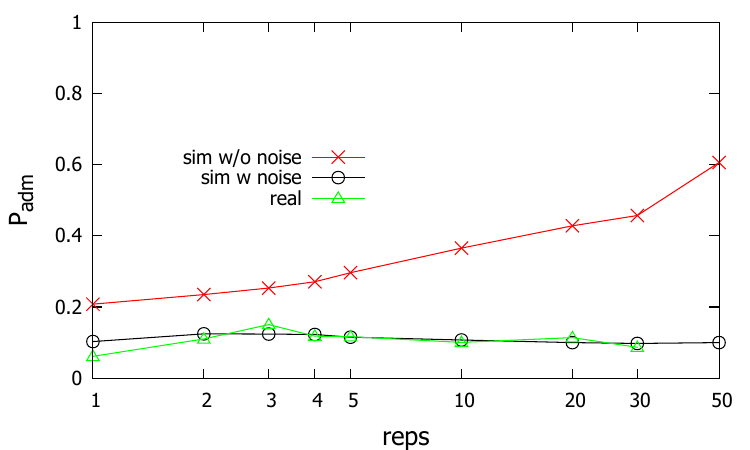}
	\caption{Probability $P_{adm}$ with 8 qubits.}
	\label{fig:confronto_8qubit_qaoa_probSuccess}
\end{figure}

\begin{figure}[!tb]
	\centering
	\includegraphics[width=0.95\columnwidth, trim=0cm 0.5cm 0cm 0cm]{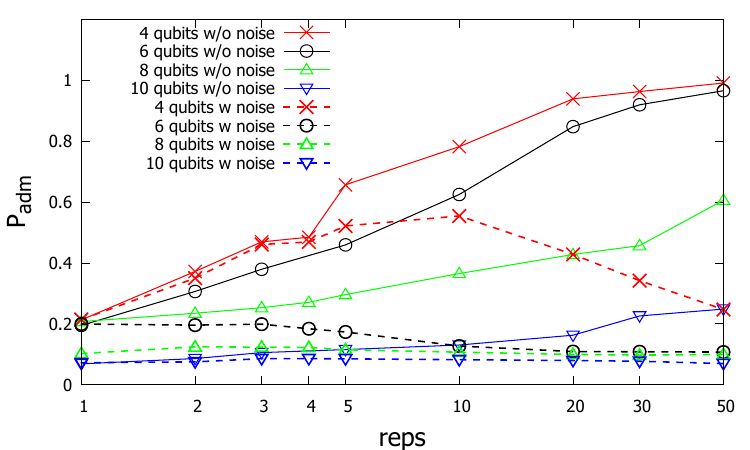}
	\caption{Probability $P_{adm}$ for different  numbers of qubits, with noiseless and noisy simulations.}
	\label{fig:confronto_alCrescereDeiQubit_qaoa_sommaProbabilitySuccessNoiseNoNoise}
\end{figure}

In problems with more qubits, not all the admissible solutions are also optimal, as it was the case with 4 qubits. Therefore, for the 6-qubit problem, we report both the $P_{best}$ and $P_{adm}$ solutions, in Figure \ref{fig:confronto_6qubit_qaoa_probfvalEsatta} and \ref{fig:confronto_6qubit_qaoa_probSuccess}, respectively. The analogous results for 8 qubits are in Figures \ref{fig:confronto_8qubit_qaoa_probfvalEsatta} and \ref{fig:confronto_8qubit_qaoa_probSuccess}. When increasing the problem size, we see that the probabilities decrease, but in noiseless simulations are still large enough to be able to achieve the optimal solution on a large fraction of the 4096 shots. For example, with 8 qubits and 50 repetitions, about $60\%$ of the 4096 shots provide an admissible solution and about $8\%$ provide the best solution. However, the impact of noise is heavier and prevents using an adequate number of repetitions, which would be required to obtain results of good quality.
Figure \ref{fig:confronto_alCrescereDeiQubit_qaoa_sommaProbabilitySuccessNoiseNoNoise} gives a direct comparison among the values of $P_{adm}$ obtained, with noiseless and noisy simulations, with the number of qubits ranging between 4 and 10.

It is expected that the in the next few years the noise of quantum computers will be considerably reduced, and also that more efficient noise-reduction algorithms will be available, but it is difficult to predict if this will be sufficient when increasing the problem size. Therefore, it is useful to investigate other techniques that can improve performance. In the following, we discuss the improvements achievable with Recursive QAOA.

\subsection{Performance of Recursive QAOA}
\label{secRecursive}

\begin{figure}[!tb]
	\centering
	\includegraphics[width=0.95\columnwidth, trim=0cm 0.5cm 0cm 0cm]{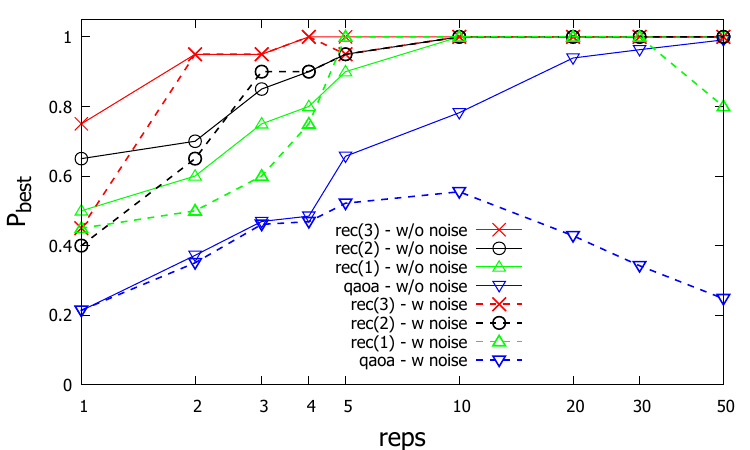}
	\caption{Probability $P_{best}$ with 4 qubits, with QAOA and recursive QAOA (denoted as ``rec(x)'', where $x$ is the value of $num\_min\_var$), with and without noise.}
	\label{fig:confronto_4qubit_recursive_probfvalEsatta_NoiseNoNoise}
\end{figure}

\begin{figure}[!tb]
	\centering
	\includegraphics[width=0.95\columnwidth, trim=0cm 0.5cm 0cm 0cm]{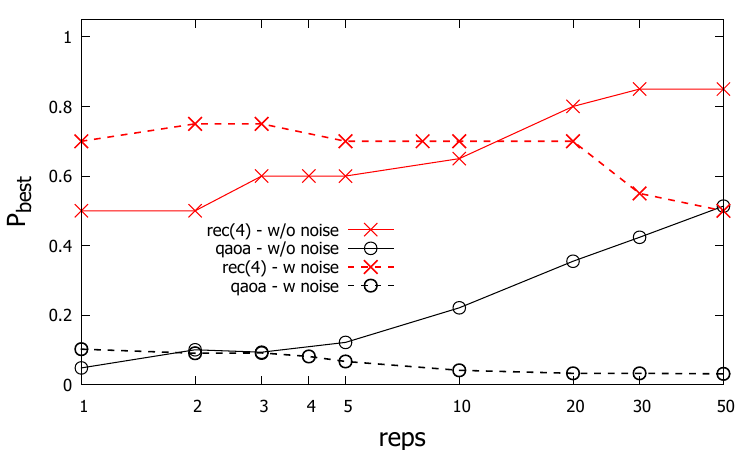}
	\caption{Probability $P_{best}$ with 6 qubits, with QAOA and recursive QAOA (denoted as ``rec(x)'', where $x$ is the value of $num\_min\_var$), with and without noise.}
	\label{fig:confronto_6qubit_recursive_probfvalEsatta_NoiseNoNoise}
\end{figure}

\begin{figure}[!tb]
	\centering
	\includegraphics[width=0.95\columnwidth, trim=0cm 0.5cm 0cm 0cm]{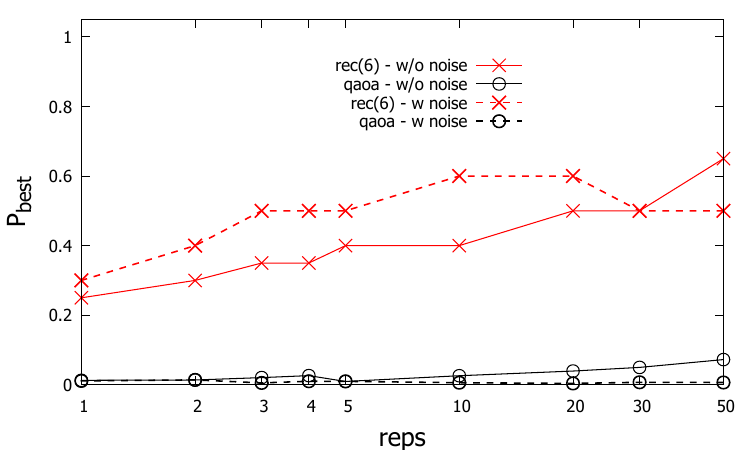}
	\caption{Probability $P_{best}$ with 8 qubits, with QAOA and recursive QAOA (denoted as ``rec(x)'', where $x$ is the value of $num\_min\_var$), with and without noise.}
	\label{fig:confronto_8qubit_recursive_probfvalEsatta_NoiseNoNoise}
\end{figure}

The first important result achieved with Recursive QAOA is that this variant was always able to provide an admissible solution, in all the scenarios considered in this paper. This means that the problem size reduction, made in the first stage of the algorithm, does not exclude at least some of the admissible solutions, which are then returned by the final classical step. Therefore, the probability $P_{adm}$ is always equal to 1.0. The performances of Recursive QAOA in terms of $P_{best}$ are also very interesting. Figure \ref{fig:confronto_4qubit_recursive_probfvalEsatta_NoiseNoNoise} compares the success probability with 4 qubits, obtained with ``regular'' QAOA and with Recursive QAOA, in noiseless and noisy simulations. We can observe that the values of $P_{best}$ are always larger with Recursive QAOA, and approach the value 1.0 when increasing the number of repetitions. Moreover, we notice that the performances are better when the value of  \textit{num\_min\_var} is larger, i.e., when fewer variables are eliminated in the first stage of the algorithm. This confirms the discussion made in Section \ref{secIntroQAOA}. Analogous comparisons are made in Figures \ref{fig:confronto_6qubit_recursive_probfvalEsatta_NoiseNoNoise} and \ref{fig:confronto_8qubit_recursive_probfvalEsatta_NoiseNoNoise} in the scenarios with 6 and 8 qubits, respectively. In these figures, for the sake of readability, results of Recursive QAOA are only reported with the value of \textit{num\_min\_var} equal to the number of qubits minus 2, meaning that two variables are eliminated before the classical optimization step. A further interesting remark is that not only Recursive QAOA performs better than regular QAOA, but it is also less sensitive to noise: results obtained with and without noise are closer to one another with Recursive QAOA than with regular QAOA. 

\begin{figure}[!tb]
	\centering
	\includegraphics[width=0.95\columnwidth, trim=0cm 0.5cm 0cm 0cm]{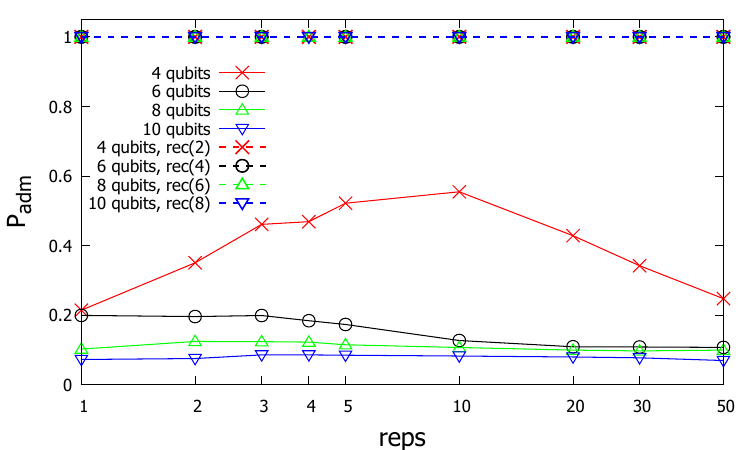}
	\caption{Probability $P_{adm}$ for different  numbers of qubits, with QAOA and recursive QAOA, in the presence of noise.}
	\label{fig:confronto_alCrescereDeiQubit_recursive_Noise_sommaProbabilitySuccess}
\end{figure}

Figure \ref{fig:confronto_alCrescereDeiQubit_recursive_Noise_sommaProbabilitySuccess} offers an overview of the performances of noisy simulations, in terms of $P_{adm},$ when varying the number of qubits. The figure highlights the ability of Recursive QAOA of finding an admissible solution in any considered scenario, while the performance of regular QAOA strongly depends on the number of qubits and on the number of repetitions.

\subsection{Scalability}
\label{secScalability}

So far, we have seen that the Recursive QAOA algorithm is able to find the optimal solution, or at least an admissible one, with a large probability, and that the noise seems to have a limited impact on its performance.
It is worth noting that the use of a larger number of qubits is impracticable on a simulator, due to the exponential explosion of the required memory structures. Indeed, a quantum register with $n$ qubits lives in a Hilbert space with dimensionality $2^n$, and the operators -- both the Hamiltonian Ising operator and the unitary operators that represent the quantum gates  -- are represented as $2^n$ x $2^n$ matrices. Hence, the assessment of large problems requires real quantum hardware, since the operators are directly executed in parallel on the qubits, and the amount of memory scales linearly -- not exponentially -- with the size of the problem. In the near future, more powerful quantum computers will be made available to the research community, and it will be possible to assess the performances of the algorithm for larger-size problems.

\begin{figure}[!tb]
	\centering
	\includegraphics[width=0.95\columnwidth, trim=0cm 0.5cm 0cm 0cm]{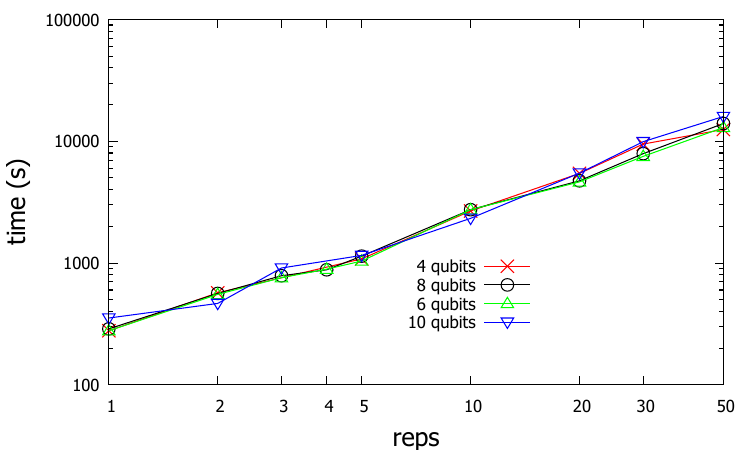}
	\caption{Computing time on the real platform \textit{ibmq\_montreal}.}
	\label{fig:confronto_alCrescereDeiQubit_qaoa_time_real}
\end{figure}

\begin{figure}[!tb]
	\centering
	\includegraphics[width=0.95\columnwidth, trim=0cm 0.5cm 0cm 0cm]{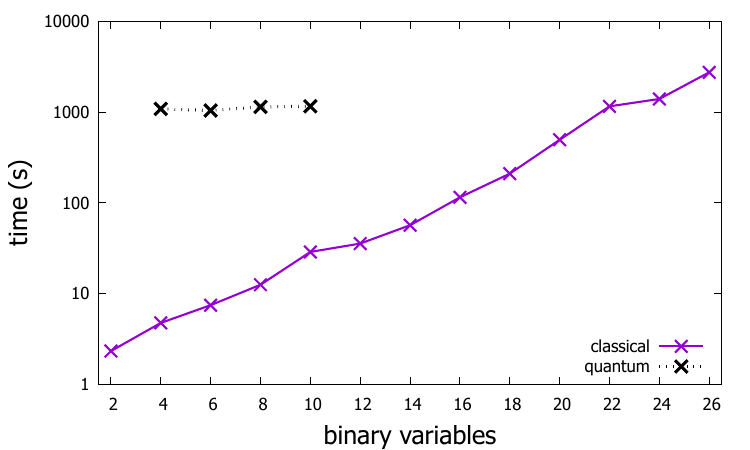}
	\caption{\blue{Execution time of the classical CPLEX solver and QAOA (with $reps=5$) vs. the number of binary variables.}}
	\label{fig:tempi_risoluzione_classica}
\end{figure}

Figure \ref{fig:confronto_alCrescereDeiQubit_qaoa_time_real} and 
Figure \ref{fig:tempi_risoluzione_classica} are helpful to assess
the scalability in terms of the computing time. \blue{Specifically, Figure \ref{fig:confronto_alCrescereDeiQubit_qaoa_time_real} reports, in log-log scale, the execution time of QAOA on the real quantum machine \textit{ibmq\_montreal} versus the number of repetitions.} The time increases linearly with the number of repetitions, as expected, since each repetition needs the execution of a fixed number of quantum gates. More significantly, the time is almost independent of the number of qubits, which confirms the potential scalability of the quantum algorithm.
\blue{Figure \ref{fig:tempi_risoluzione_classica} compares, in a semi-log plot, the computing times needed by QAOA vs. those experienced with the well-known classical CPLEX solver\footnote{Qiskit CplexOptimizer: \url{https://qiskit.org/documentation/optimization/stubs/qiskit_optimization.algorithms.CplexOptimizer.html}}, for different numbers of binary variables. 
The results show that the computing time obtained with CPLEX increases exponentially with the number of binary variables, while the quantum computing time is almost constant.
Even if the numerical values of computing time are larger than those obtained with classical computing, the dramatic difference between the two trends confirms the hope for a significant speed-up of quantum computing algorithms for large problems.}


\subsection{Perspective of Quantum Computing for Energy Scheduling}



\blue{Also in light of the experimental results, we can now try to make the point on the possible exploitation of variational quantum algorithms (VQAs) for the resolution of the prosumer problem and energy scheduling problems in general. As we have seen, linear programming problems
can be reformulated as Ising problems, thus enabling the exploitation of QAOA. More complex and non-linear problems
can be tackled with other VQA algorithms, which, as mentioned in the introductory section, are seen as the quantum analog of neural networks \cite{Chang2020OnHQ, VQA}: in this case, the objective is the minimization of a cost function, computed starting from the results of quantum measurements, and the approach is still hybrid and variational, with a parameterized quantum circuit that is optimized with classical methods, e.g., gradient descent. The advantage of QAOA is that the theory predicts that the process converges to the optimal result (the ground state) when increasing the number of repetitions, which is not guaranteed in other VQAs.}

\blue{The main expected benefits are the same for QAOA and other VQAs: (i) better scalability with respect to classical algorithms and (ii) efficient and natural management of the uncertainty. The hope for better scalability originates from the possibility of exploring an exponential state space and from the use of quantum gates that operate in parallel on all of the basis states, which are also exponential in number. The results shown in Section \ref{secScalability} are encouraging in this respect. With regard to the management of the uncertainty, we have seen that QAOA returns a set of results, which can be selected ``a posteriori'' after some possible variations of the conditions, and this holds also for other VQAs. Moreover, in presence of slight changes in the input data, variational algorithms can be re-executed setting the initial values of the parameters to the values obtained after the previous optimizations, which can be expected to shorten the computation. 
Indeed, whenever the solution corresponds to the ground state of an Hamiltonian, as in QAOA, it tends to follow it adiabatically, remaining there even after small changes occur in the environmental conditions.}

\blue{Putting the use of quantum computing methods into perspective, there are also important challenges to be faced. Firstly, the noise and decoherence of available quantum hardware, whose effect increases with the depth of the circuit, are already significant, despite the fact that we addressed small size problems. Any reliable evaluation can be done only after more efficient and larger hardware will become available. Secondly, while we have seen that the complexity of the QAOA circuit is polynomial, there is no guarantee about the number of iterations. This problem derives from the phenomenon referred to as ``barren plateaux'' \cite{barren}, i.e., the magnitude of partial derivatives tends to vanish
with the system size, which can hamper the discovery of a clear path for the optimization of the parameter values.
A further challenge on the applications on VQA is that the number of measurements required for the estimation of the cost function can become excessive, thus hindering the efficiency of the process \cite{VQA}. Despite these though challenges, the potential benefits are important, and the research work in the next few years will determine whether quantum algorithms are a viable alternative for the solution of energy scheduling problems.}





\section{Conclusions}
\label{secConclusions}
This paper focuses on the solution of the prosumer problem with a hybrid classical-quantum computing approach. We have outlined how this NP-hard problem can be transformed into the problem of finding the ground state of a Hamiltonian operator, which is the kind of problem that can be solved by the QAOA and Recursive QAOA algorithms.
We have tested the performance of these algorithms
in finding the best and admissible solutions, by using both real and simulated resources. Moreover, we have compared the results of QAOA to those of Recursive QAOA \cite{QAOA-recursive}, showing that the latter significantly over-performs the former in addressing the prosumer problem, both with noisy and noiseless quantum hardware. Finally, despite performing our tests with a limited number of qubits, we have been able to inspect the scalability of the algorithms:
we have checked that the quantum execution time
does not depend on the number of binary variables and
increases almost linearly with the requested accuracy. This suggests that, as the problem size increases, a quantum approach is expected to be technologically favorable in the long term.




\section*{Acknowledgments}
This work was partially funded by the Italian MUR Ministry under the project PNRR National Centre on HPC, Big Data and Quantum Computing, PUN: B93C22000620006, and by
European Union - NextGenerationEU - National Recovery and Resilience Plan (Piano Nazionale di Ripresa e Resilienza, PNRR) - Project: “SoBigData.it - Strengthening the Italian RI for Social Mining and Big Data Analytics” - Prot. IR0000013 - Avviso n. 3264 del 28/12/2021.


\bibliographystyle{IEEEtran}
\bibliography{bibliography}


\section*{Biography Section}

 



\begin{IEEEbiography}[{\includegraphics[width=1in,height=1.25in,clip,keepaspectratio]{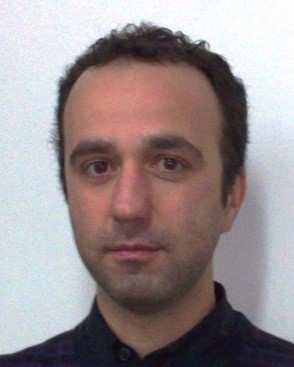}}]{Carlo Mastroianni}
received the Laurea degree and the PhD degree in computer engineering from the University of Calabria, Italy, in 1995 and 1999, respectively.
He is a Director of Research with ICAR-CNR, Rende, Italy. He has coauthored over 100 papers published in international journals and conference proceedings. His current research interests include distributed computing, internet of things, cloud computing, bio-inspired algorithms, smart grids and quantum computing.
\end{IEEEbiography}

\begin{IEEEbiography}[{\includegraphics[width=1in,height=1.25in,clip,keepaspectratio]{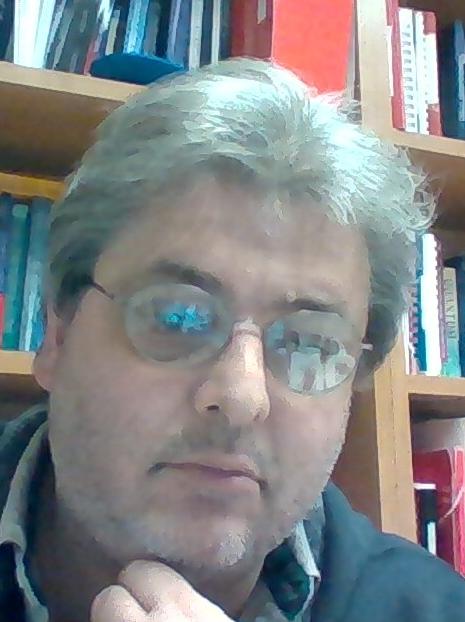}}]{Francesco Plastina} received his Ph.D. in Physics at University of Calabria in 2000, where he is now associate professor in theoretical physics. He is coauthor of more than 80 research papers, mainly focusing on quantum information theory, quantum thermodynamics, quantum coherence and correlations, and open quantum systems.  
\end{IEEEbiography}

\begin{IEEEbiography}[{\includegraphics[width=1in,height=1.25in,clip,keepaspectratio]{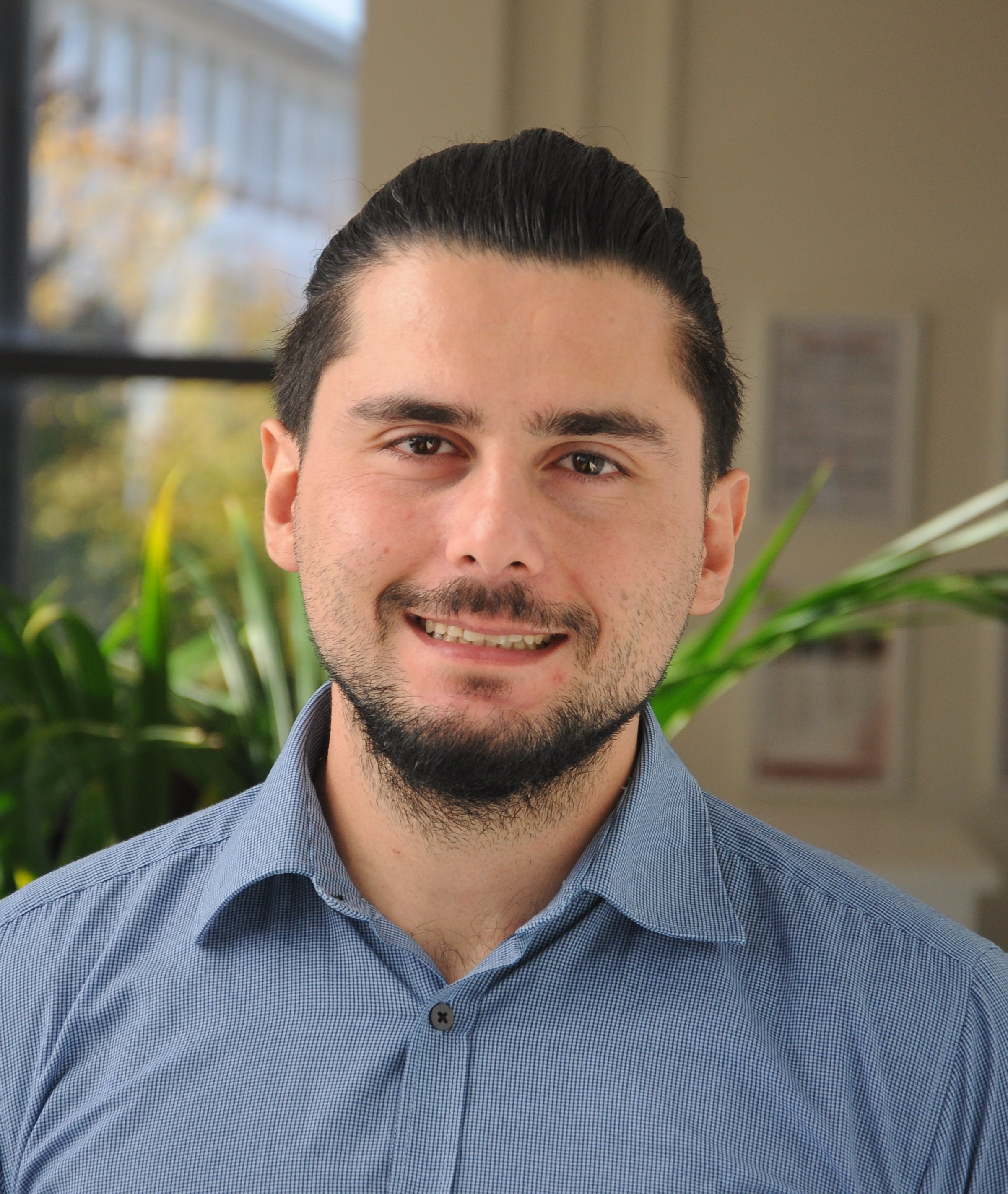}}]{Luigi Scarcello}
received his master's degree in Energy Engineering at the University of Calabria, Italy, in 2015.
He got his PhD title in Science and Engineering of Environment, Construction and Energy, in 2019.
He is a researcher at the Institute of High Performance Computing and Networking, National Research
Council, Rende. His many research activities run on optimization models and enabling IoT technologies for energy management
in smart grids and energy communities.\end{IEEEbiography}

\begin{IEEEbiography}[{\includegraphics[width=1in,height=1.25in,clip,keepaspectratio]{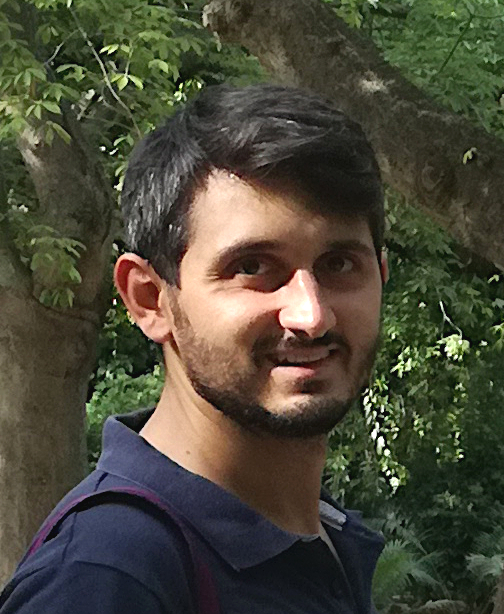}}]{Jacopo Settino}
received his master's degree in Physics at the University of Calabria, Italy, in 2015.
He got his PhD Doctor Europaeus title in Physical, Chemical and Materials Sciences and Technologies, in 2019.
After a two-year post-doc at SPIN-CNR he is a researcher at the Physics Department of University of  Calabria, Rende. His many research activities run on quantum algorithms, topological superconductors, disordered and interacting quantum systems.
\end{IEEEbiography}

\begin{IEEEbiography}[{\includegraphics[width=1in,clip,keepaspectratio]{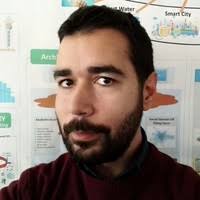}}]{Andrea Vinci}
received a PhD in system engineering and computer science from the University of Calabria, Rende, Italy. He is a Researcher with ICAR-CNR, Rende, Italy,
since 2012. His research mainly focuses on the Internet of Things and cyber-physical systems. He has authored or co-authored researches on the definitions of platforms and methodologies for the design and implementation of cyber-physical systems, and on distributed algorithms for the efficient control of urban and building infrastructures based on artificial and swarm intelligence.
\end{IEEEbiography}



\vfill

\end{document}